\documentstyle[aps,manuscript,epsf]{revtex}  
\include{psfig}
\title{Statistical Physics of Unzipping DNA}
\author{David R. Nelson}
\address{Lyman Laboratory of Physics, Harvard University, 
Cambridge, MA 02138.\\
E-mail: nelson@cmt.harvard.edu}
\date{\today}
\begin{document}
\maketitle
\begin{abstract}
The denaturation of double-stranded DNA as function of 
force and temperature is discussed. At room temperature, 
sequence heterogeneity dominates the physics of 
single molecule force-extension curves starting 
about 7 piconewtons of below a $\sim$15 $pN$ unzipping
transition. The dynamics of the unzipping fork exhibits  
anomalous drift and diffusion in a similar range above this
transition. Energy barriers 
near the transition scale as the square root 
of the genome size. Recent observations of   
jumps and plateaus in the unzipping of lambda phage DNA at 
constant force are 
consistent with these predictions.
\bigskip
\bigskip

\end{abstract}
\narrowtext

\section{Single Molecule Biophysics Experiments}

The past decade has seen a revolution in biophysics,
 due to 
exquisitely sensitive experiments \cite{bustamante}
which probe ingredients of the ``central dogma'' 
of molecular biology \cite{crick} at the level of 
{\it individual} molecules. Although exceptions exist, 
the central dogma states that biological 
information first flows from 
DNA to messenger RNA via transcription mediated by a 
RNA polymerase. Information is then transferred from mRNA 
to proteins via a ribosome-mediated translation process.

Among the  efforts to probe the basic constituents further:
(a)~Proteins anchored to a microscope slide in their 
biologically relevant folded state have been teased apart 
with atomic force probes \cite{rief} attached using, e.g., 
biotin-streptavidin linkages; 
(b)~individual DNA's linked to magnetic beads have been 
stretched and twisted by small magnetic field 
gradients, allowing studies of both supercoiling 
and twist-induced denaturation \cite{strick}; and  
(c)~the reversible unfolding of single RNA molecules (with 
beads attached to both ends) has been studied using laser 
tweezers with a feedback loop to generate a constant force 
$F$ \cite{lip}. In this later experiment, the fluctuating 
displacements of a $\sim$25 base-pair RNA hairpin evolve from a 
predominantly closed state (with rare opening events) 
at small force to a predominantly open state (with rare closing events) 
at larger forces. At an intermediate 
force of $F_c \approx 14$ piconewtons 
$(pN)$, the hairpin spends approximately equal time in both 
configurations, suggestive of a first-order phase 
transition between the open and closed states in the 
``thermodynamic limit'' of a very large hairpin.
\begin{figure}[h]
\begin{center}
\centerline{\psfig{figure=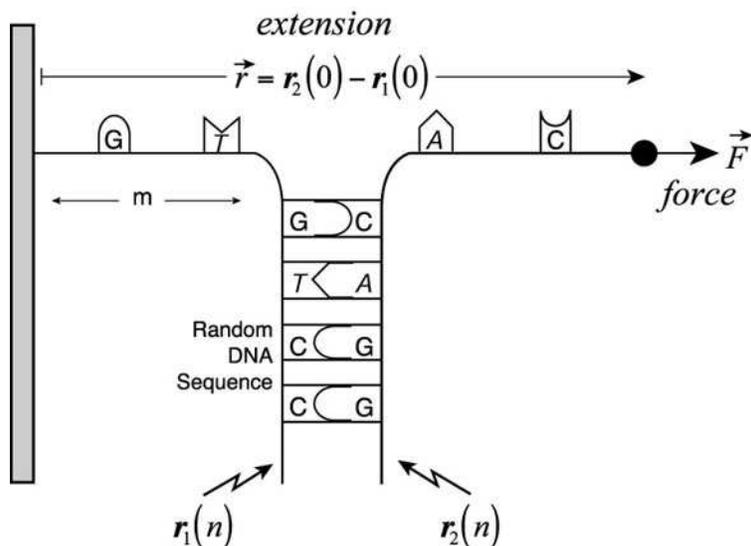}}
\parbox{6.5in}
{\caption{Schematic of DNA unzipping experiments. The sugar-phosphate
backbones, with complementary base pairing, are represented by the 
functions ${\bf r}_1(n)$ and ${\bf r}_2(n)$, where $n$ 
indexes base pairs. Experiments can be carried out either at 
constant extension ${\bf r}_1(0)-{\bf r}_2(0)$ or at constant force $\vec{F}$.}
}
\medskip
\end{center}
\end{figure}
DNA force denaturation experiments related to the 
subject of this review have been carried by 
Bockelmann {\it et al.} \cite{bock,esse,bocktho} 
(see Fig. 1).  In this pioneering work, the 
48,502 base pairs of the virus phage lambda with 
one strand attached to a microscope slide were 
unzipped by a microneedle attached to the other 
strand under conditions which produced a constant 
rate of unzipping at a fixed temperature. Reasonable 
agreement with these experiments was obtained by direct 
numerical evaluation of the equilibrium statistical 
sums associated with a given average degree of 
unzipping. If the rate of unzipping is slow, such 
experiments effectively probe the physics at constant 
extension. The average force under these conditions 
locks in at about 14--15 piconewtons. The 
time-dependent fluctuations of this force as the 
unzipping proceeds provide information about the 
particular sequence of G:C and  A:T base 
pairs being torn apart \cite{bock,esse,bocktho}.
The full phase diagram of DNA in the 
force-temperature plane is shown in Fig. 2 \cite{lubens,bhat,sebastian,cocco,moren}. 
The thermal DNA melting transition at zero force 
(see Sec. 2) is characterized by diverging 
length scales associated with denaturation 
bubbles. When $F \neq 0$, the native duplex  
DNA denatures into an unzipped state via a 
first-order phase transition (see Sec. 3). 
Under conditions of a slow, constant rate of 
extension \cite{bock,esse,bocktho}, the average force 
{\it adjusts} so that the system sits on the  
heavy first-order transition line $F_c(T)$. At 
any stage in the unzipping process, the unzipping 
fork separating the native and unzipped states 
is like the meniscus dividing, say, liquid and 
gas phases at constant volume at a bulk 
first-order phase transition.
\begin{figure}[h]
\begin{center}
\centerline{\psfig{figure=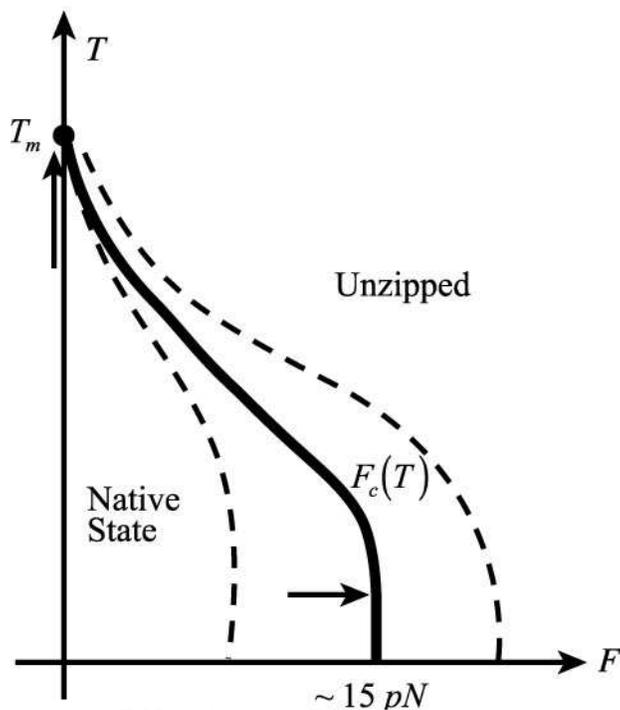}}
\parbox{6.5in}
{\caption{Phase diagram for DNA denaturation in the force-temperature 
plane. Conventional thermal denaturation occurs at zero force as $T$
approaches the melting temperature $T_m$. Force-induced denaturation, 
or unzipping occurs on the path indicated by the horizontal arrow.
Sequence heterogeneity dominates the physics between the two 
dashed crossover lines.}
}
\end{center}
\end{figure}
\begin{figure}[h]
\begin{center}
\centerline{\psfig{figure=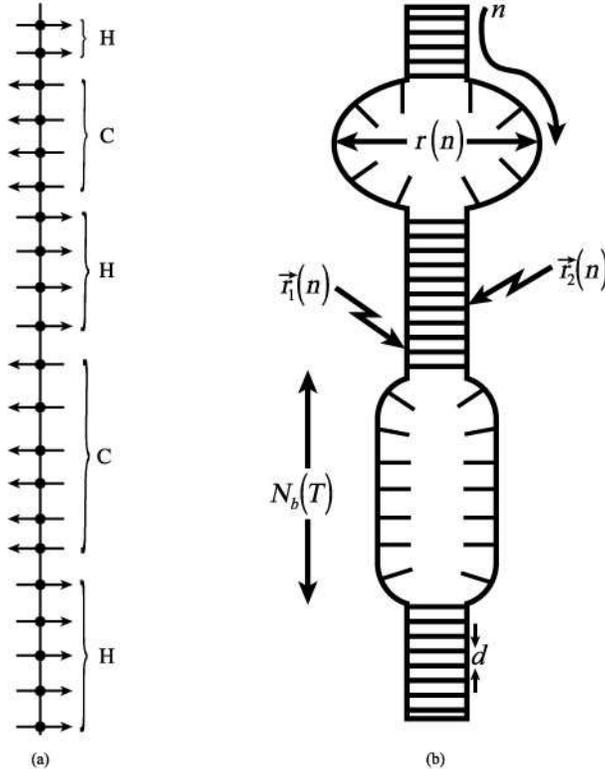}}
\parbox{6.5in}
{\caption{Models of thermal denaturation. (a) Ising model, where right-pointing
spins represent hydrogen-bonded ``helix'' segments, and left-pointing spins
represent denatured ``coil'' segments. (b) More sophisticated polymer model 
of denaturation whose statistical mechanics is mapped onto a problem in 
three-dimensional quantum mechanics. $N_b(T)$ is the number of base 
pairs associated with a typical denatured ``coil'' segment.}
}
\end{center}
\end{figure}

Although running unzipping experiments in a 
two-phase region associated with the line $F_c(T)$ is 
well-suited for possible DNA sequencing 
applications \cite{bock,esse,bocktho}, the full force-temperature 
phase diagram of Fig. 2 is interesting for a number 
of reasons. First, understanding DNA denaturation 
under conditions of constant force 
(instead of constant extension) could 
provide insights into the complicated process 
by which DNA replicates during bacterial cell division 
\cite{alberts}. Second, statistical mechanics in 
ensembles with {\it intensive} variables like the 
force held fixed is often more tractable than when 
the control parameter is a conjugate variable like 
the extension. Precise analytic predications are 
indeed possible in the vicinity of $F_c$, even when 
sequence disorder is present \cite{lubens,lubensnelson}. Third, 
experiments which use force 
as a control parameter can probe interesting 
behavior of the unzipping fork for $F < F_c(T)$ 
(analogous to ``wetting'' phenomena in bulk phase 
transitions \cite{gennes}) and anomalous fork 
dynamics for $F > F_c(T)$. Because of significantly 
different bonding strengths of the nucleotide 
pairs  A:T (two hydrogen bonds) and  G:C 
(three hydrogen bonds), heterogeneity dominates 
unzipping over a large region of the phase diagram,
for $\frac{1}{2} F_c\lesssim F\lesssim \frac{3}{2} F_c$, 
inside the region bounded by the dashed lines in Fig.~2.

\section{Thermal Denaturation of DNA}

In this Section we discuss the thermal denaturation 
of DNA at zero force (vertical arrow in Fig. 2). 
This transition plays an important role in the 
polymerase chain reaction (PCR), where it is applied 
cyclically to amplify minute amounts of DNA. 
The basic physical ideas driving thermal denaturation 
(worked out in the 1960s) 
are described in Refs. \cite{polan,wiegel,gros}. The simplest 
models invoke the Ising-like description of DNA shown 
in Fig. 3a, where right-facing spins represent 
bonded nucleotides in a closed 
``helix'' state and left-facing spins the open 
``coil'' state. The magnetic field 
of the equivalent Ising model represents a 
(temperature-dependent) 
nucleotide bonding free energy and the Ising 
exchange energy determines an ``initiation factor'' 
at the boundaries between the 
helix and coil regions. The statistical 
mechanics of this one-dimensional Ising model 
can be described entirely in terms of 
kink excitations associated with transitions 
between domains of up and down spins, 
which here represent helix/coil boundaries. 
Rapid variations in DNA from a helix-dominated 
to a coil-dominated state are predicted 
by such models \cite{gros}, corresponding to the 
magnetic field changing sign in the 
equivalent Ising system. However, 
it is well-known that the 1d Ising model 
does not have a real 
phase transition at finite temperature, 
basically because the finite energy 
cost of kinks at zero external 
field is overwhelmed by the 
entropy gained by creating them 
[20].  An improved theory 
results when the three-dimensional 
entropy of wandering loops 
in the coil sections (see Fig. 3b) 
is taken into account. This can
be done by approximating the loops 
as ideal random walks \cite{polan},  
or improving this estimate by taking 
self-avoidance of the loops 
into account \cite{fisher}. The effect on the 
statistical mechanics is 
to produce a long range interaction 
which binds kinks together and 
leads to a continuous finite temperature 
phase transition \cite{polan,wiegel,gros}.
Recent work by Kafri {\it et al.} incorporates 
additional self-avoidance 
between coil sections and the rest of the 
DNA chain (important for 
large molecules) and argues that 
the thermal denaturation transition then becomes first order
\cite{kafri}.

Heterogeneity of the bonding between 
base pairs (G:C bonds are stronger than 
A:T ones) is neglected in the treatments above. 
Simple helix-coil descriptions with 
heterogeneity are governed by the statistical 
mechanics of the 1d random field 
Ising model. A reasonable guess for 
more realistic models with heterogeneity is 
that while the transition may be less sharp 
(weakly bound regions melt  before more tightly 
bound ones), a real finite temperature phase 
transition survives. Lyubchenko {\it et al.} \cite{lyub} 
have calculated numerically the effect of 
sequence heterogeneity on the DNA differential 
melting curves for the 5375 nucleotides 
of the virus $\phi X174$. For a recent theory of 
thermal denaturation of heterogeneous DNA, 
see Ref. \cite{tang}.

In the remainder of this section, we discuss 
how thermal denaturation of ``homogeneous'' 
DNA with identical base pairs
 can {\it also} be understood 
by mapping the statistical mechanics 
onto the delocalization of a quantum 
mechanical particle in three dimensions. 
This is the approach we will 
generalize in Sec. 3 to study 
force-induced naturation.

Referring to Fig. 3b, we parameterize the 
positions of the two (antiparallel) 
sugar-phosphate backbones of DNA in three dimensions 
by the functions ${\bf r}_1(n)$ and 
${\bf r}_2(n)$,  where $n$ is an integer 
indexing $N$ base pairs $(0 \leq n \leq N)$. We 
model each strand separately as a Gaussian random 
coil \cite{doi} . For example, considering strand 1 
in isolation, the probability of a particular 
polymer conformation ${\bf r}_1(n)$ 
is proportional to $e^{-F_1[{\bf r}_1(n)]/T}$
(we use units such that $k_B=1$), where 
\begin{equation}
F_1[{\bf r}_1(n)]=
\frac{1}{2} K\int_{0}^{Nd}
\left(\frac{d{\bf r}_1(s)}
{ds}\right)^2\;\;ds.
\end{equation}
Here we have used a convenient continuum notation 
as shorthand for a discrete sum over 
$n$ and replaced $n$ by the 
arclength $s = nd$, where $d$ is
the spacing between nucleotide monomers. 
To relate the spring constant 
$K$ to the single strand persistence 
length, use (1) to evaluate
$\langle|\vec{r}_1(Nd)-\vec{r}_1(0)|^2\rangle$, with the result 
\begin{equation}
\langle | \vec{r}_1(Nd)-\vec{r}_1(0)|^2\rangle
=\frac{3TNd}{K}.
\end{equation}
From the definition of persistance length $\ell$, namely
$\langle|\vec{r}_1(Nd)-\vec{r}_1(0)|^2\rangle\equiv
\ell Nd$, we have \cite{nelson}
\begin{equation}
K=\frac{3T}{\ell},
\end{equation}
where $\ell\approx 10$\AA \ for single-stranded DNA at 
physiological temperatures. For a more microscopic
``worm-like chain'' model with bending rigidity $\kappa$ (assuming
a small Debye screening length), we 
have \cite{degennes,marko}
\begin{equation}
\ell\approx\frac{2\kappa}{T}.
\end{equation}
Here, and in what follows, we  use boldface vectors
${\bf r}(s)$ to denote entire polymer configurations, and conventional vectors such
as $\vec{r}(Nd)$ to denote the position of a polymer endpoint.

To describe a DNA duplex, we bring the two strands together, 
neglect torsional rigidity and the helical nature of the bonding 
\cite{marko}, 
and write the total free energy as  
\begin{eqnarray}
F[{\bf r}_1(s),{\bf r}_2(s)] & =
 \frac{K}{2}
\int_{0}^{Nd}
\left[\left(
\frac{d{\bf r}_1(s)}{ds}\right)^2 +
\left(\frac{d{\bf r}_2(s)}{ds}\right)^2\right]ds\nonumber \\
&\quad +
\int_{0}^{Nd}
U[{\bf r}_1(s)-
{\bf r}_2(s)]\;\;ds,
\end{eqnarray} 
where
$U[{\bf r}_1(s)-
{\bf r}_2(s)]$ is a potential which binds the nucleotides in different strands 
together. We assume that the complimentarity of the nucleotide sequences is 
sufficient to bring the strands into registry, but will otherwise neglect 
the effect of sequence disorder in the remainder of this section. If we now
pass to sum and difference variables.
\begin{equation}
{\bf R}(s)=\frac{1}{2}[{\bf r}_1(s)
+{\bf r}_2(s)],
\end{equation}
\begin{equation}
{\bf r}(s) = {\bf r}_2(s)-
{\bf r}_1(s),
\end{equation}
the coarse-grained free energy decouples,
\begin{eqnarray}
F[{\bf R}(s),{\bf r} (s)] & =
K \int_0^{Nd}
ds \left( \frac{d{\bf R}}{ds}\right)^2+
\frac{1}{2} g\int_0^{Nd} 
ds \left( \frac{d{\bf r}}{ds}\right)^2\nonumber \\
&\quad  
+\int_0^{Nd} ds\; U[{\bf r} (s)],
\end{eqnarray}
where $g=K/2$.

The partition function associated with (8)  is a path integral
of $e^{-F({\bf R}, {\bf r} )/T}$ over the functions ${\bf R}(s)$ and ${\bf r}(s)$.
The part associated with ${\bf R}(s)$ is simply the partition sum of 
an unconstrained Gaussian coil. The remaining functional integral
over ${\bf r}(s)$ contains a denaturation transition. To pursue this 
point, we assume for simplicity that ${\bf r}(s=0)=\vec{0}$ and 
${\bf r}(s=Nd)=\vec{r}$ (closed boundary conditions at one end, 
open at the other), and evaluate
\begin{equation}
Z(\vec{0},\vec {r}; N)=
\int_{{\bf r} (0)=\vec{0}}^{ {\bf r} (Nd)=\vec{r}}
{\cal D}r \exp\left[
-\frac{g}{2T}\int_0^{Nd}
\left(\frac{d {\bf r} }{ds}\right)^2
ds-\frac{1}{T}\int_0^{Nd}
U[{\bf r}(s)]\right]\;ds.
\end{equation}
As in polymer adsorption problems, it is helpful to view Eq. (9) as a 
Feynman path integral for a quantum mechanical particle in imaginary
time 
\cite{degennes}. 
Indeed, this partition function can be rewritten as a 
quantum mechanical matrix element \cite{nelsonbook},
\begin{equation}
Z(\vec{0},\vec{r}; N)=
\langle \vec{r}|e^{-\hat{H} Nd/T}|\vec{0}\rangle,
\end{equation}
where the effective quantum Hamiltonian is
\begin{equation}
\hat{H}=\frac{-T^2}{2g}
\nabla^2+U(\vec{r})
\end{equation}
and $|0\rangle$ and $\langle\vec{r}|$ are respectively $ket$ and $bra$ 
vectors localized at $\vec{0}$ and $\vec{r}$. Note that 
temperature plays the role of $\hbar$ and $g=\frac{1}{2}K=\frac{3T}{2\ell}$ 
represents the mass of a fictitious quantum mechanical particle in a 
potential $U(\vec{r})$. In the language of statistical mechanics,
we have reduced this one-dimensional problem to diagonalizing a 
transfer matrix given by $\hat{T}=e^{-\hat{H}d/T}$. We shall focus
on the particularly simple binding potential illustrated in Fig. 4, namely
\begin{eqnarray*}
U(\vec{r})=\left\{
\begin{array}{rl}
\infty ,  &  r< c\nonumber \\
-U_0 , &  c< r < b\nonumber \\
0 , & b < r, \nonumber \\ 
\end{array}
\right.\nonumber\\
\end{eqnarray*}
with $U_0=V_0/d$, where $V_0$ is the average bonding energy per nucleotide. Upon 
inserting a complete set of energy eigenfunctions $|n\rangle=
\phi_n(\vec{r})$ with eigenvalues $\epsilon_n$ into Eq. (10), we see that 
this conditional partition function may be rewritten as
\begin{equation}
Z(\vec{0},\vec{r}; N)=\sum_n
\langle\vec{r} |e^{-\hat{H}Nd/T}|n\rangle
\langle n|0\rangle= \sum_n\phi_n^*(\vec{0})
\phi_n(\vec{r})
e^{-\epsilon_n Nd/T}.
\end{equation}
\begin{figure}[h]
\begin{center}
\centerline{\psfig{figure=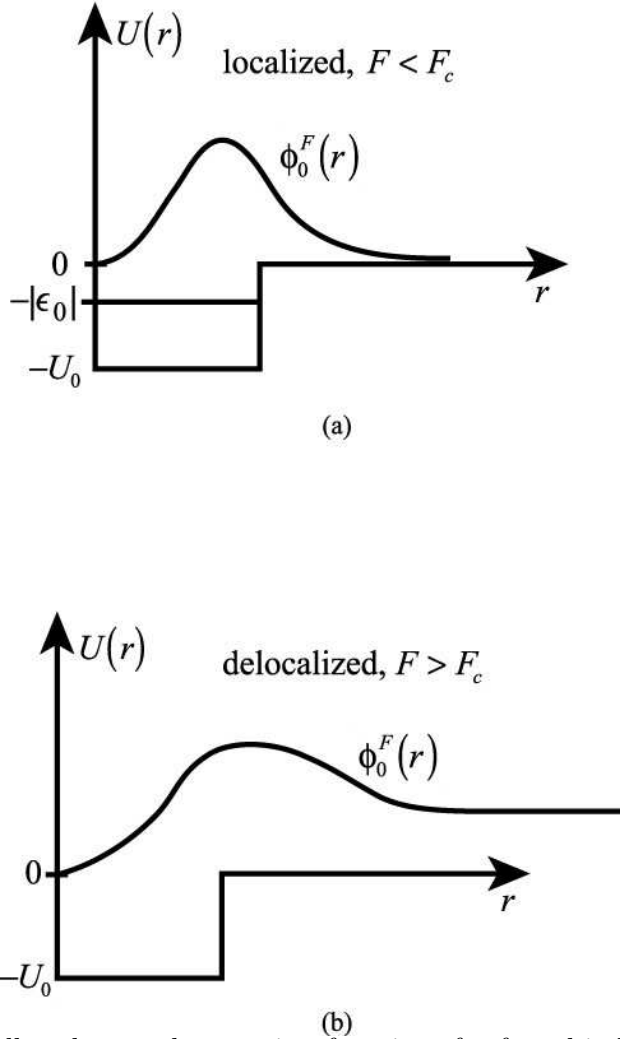}}
\parbox{6.5in}
{\caption{Potential well and ground state eigenfunctions for forced-induced 
denaturation when (a) $F<F_c$ and (b) $F>F_c$. These (right) 
eigenfunctions give the probability of a separation $\vec{r}$ for the 
unzipped ends of the DNA. For $F<F_c$, the ground state eigenfunction 
decays exponentially to zero outside the well, which models the hydrogen
bonds and stacking energies which hold base pairs together. For $F>F_c$, 
the eigenfunction tends to a nonzero constant for large $r$, and the 
two DNA strands fall apart.}
}
\end{center}
\end{figure}
The ground state dominates in the limit $N\rightarrow\infty$, and we have
\begin{equation}
Z(\vec{0},\vec{r}; N)
\approx \phi_0(\vec{0})
\phi_0(\vec{r} )
e^{-\epsilon_0 Nd/T},
\end{equation}
where the (nodeless) ground state eigenfunction can be chosen to be real.

The quantum mechanics of simple square well potentials like
Eq. (12) in three dimensions is well understood \cite{schwab}.
We focus for simplicity on the case $c<<b$ 
(small hard core diameter) but the same qualitative 
discussion applies more generally. The 
key parameter controlling the eigenfunctions and eigenvalues is the 
dimensionless ratio
\begin{equation}
Q=\frac{2gU_0b^2}{T^2}
\end{equation}
of the well depth $U_0$ to the ``zero point energy'' $\frac{T^2}
{2gb^2}$. This quantum zero point energy  represents the lost
entropy associated with a confined DNA duplex. From the 
perspective of quantum mechanics, the loop excitations shown in Fig. 3b represent 
transient excursions (in imaginary time) out of the well.
When $Q>>1$ (low temperatures), the binding energy dominates the thermal
fluctuations favoring denaturation and there are many bound (i.e., localized)
eigenstates of {\emph{$\hat H$}} beneath a continuum of extended eigenstates.
As $Q$ decreases, more and more states delocalize until only the single 
localized ground state shown in Fig. (4a) is present. Eventually,
at temperatures high enough so that $Q\le Q^*$, even the ground state
delocalizes [31]. For the simple problem discussed here $Q^*=\frac{\pi^2}{4}$
(i.e., $T_m=8 U_0b^2/\pi^2g$) [30]. However, a similar
$Q^*$ of order unity is expected for a broad class of similar
potentials. Near the transition temperature $T_m$, the ground
state energy approaches the continuum quadratically with temperature [27, 30]
\begin{equation}
\epsilon_0(T)\approx - {\rm const.} |T_m-T|^2, 
\end{equation}
where the melting temperature $T_m$ is defined by $Q
(T_m)=Q^*$. For large $r$, the ground state wave 
function decays exponentially, $\phi_0(\vec{r})\sim e^{-\kappa_0|\vec{r}|}$,
where
\begin{equation}
\kappa_0^{-1}(T)=
\sqrt{\frac{-T^2}{2g\epsilon_0(T)}}\sim \frac{1}{(T_m-T)}
\end{equation}
gives the spatial extent (in three dimensions) of the bubbles shown 
schematically in Fig. 3b. The {\it number} of base pairs $N_b$ 
in a bubble (see Fig. 3b) can be estimated by identifying $N_b$ 
with the value of $N$ needed to achieve ground
state dominance in Eq. (12): We expect that boundary effects 
become unimportant when the number of base pairs exceeds
the number in a typical bubble. The ground state will dominate once the 
number of nucleotides is larger than several bubble sizes. Since the 
energy of the first excited state for $Q\gtrsim Q^*$ is zero, we have
\begin{equation}
Z(\vec{0},\vec{r};N)\approx
\phi_0(\vec{0}) \phi_0(\vec{r})
e^{\frac{|\epsilon_0|Nd}{T}}
\left[ 1+{\cal O} \left( e^{
\frac{-|\epsilon_0|Nd}{T}} \right) \right],
\end{equation}
from which it follows that
\begin{equation}
N_b\approx
\frac{T}{d|\epsilon_0(T)|}
\sim\frac{1}{(T_m-T)^2}.
\end{equation}
We should stress that the above results are obtained in an approximation which
neglects self-avoidance effects important for large molecules close to $T_m$.
We expect, however, qualitatively similar behavior even when self-avoidance is taken
into account.
We now use similar methods to treat force-induced denaturation in Section~3.

\section{Force-induced Denaturation}

A force $\vec{F}$ applied across the two strand endpoints 
(see Fig. 1) adds an energy $-\vec{F}\cdot\vec{r}$, and Eq. (9) becomes
\begin{eqnarray}
Z(\vec{0},\vec{r}; N) & =
\int_{ {\bf r} (0)=\vec{0}}
^{ {\bf r} (Nd)=\vec{r}}
D {\bf r}(s)\exp \left[
-\frac{g}{2T}\int_0^{Nd}
\left( \frac{d {\bf r} }{ds}\right)^2 ds-\frac{1}{T}
\int_0^{Nd} U_s[{\bf r}(s)]ds \right. \nonumber \\
&\quad\left. + \frac{\vec{F}}{T}
\cdot \int_0^{Nd}
\frac{d {\bf r} }{ds} ds
\right]. 
\end{eqnarray}
We assume the attached strands of DNA can swivel freely 
to relax twist on an experimental time scale, so that we 
can neglect the helical nature of the duplex state when the 
force is applied. 
In addition to the force term (note that the lower limit does 
not contribute with our boundary conditions), we have added a subscript $s$ to the binding 
potential, $U[{\bf r}(s)]\rightarrow U_s[{\bf r} (s)]$,
to emphasize that the depth and size of potentials like that in Fig. 4
will depend on the particular base pair for heterogeneous DNA. After
taking over results for a virtually identical problem \cite{hatano} of flux line
depinning by a columnar pin in a transverse magnetic field in high 
$T_c$ superconductors, we obtain \cite{lubens,bhat}
\begin{equation}
T\frac{\partial Z(\vec{0},\vec{r}; N)}
{\partial Nd}=
\frac{-T^2}{2g}
\left( \vec{\nabla}-\frac{\vec{F}}{T}\right)
^2 Z(\vec{0},\vec{r};N)+
U_{Nd}(\vec{r})Z(\vec{0},\vec{r};N),
\end{equation}
which can be integrated (subject to the boundary condition
$Z(\vec{0},\vec{r};0)=1$) to give a result with the form of Eq. (10),
where
\begin{equation}
\hat{H}(\vec{F})=
\frac{-T^2}{2g}
\left( \vec{\nabla}-\frac{\vec{F}}{T}\right)^2+
U_{Nd}(\vec{r}).
\end{equation}

Equation (20) with an $s$-dependent potential $U_s[\vec{r}]$ can 
be studied (via a mapping onto a Burgers equation) by the methods of 
Ref. \cite{ertas}, where it arose in a study of vortex 
depinning from {\it fragmented} columnar pins. In this 
section, we neglect the $s$-dependence (letting $U_{Nd}(\vec {r})
\rightarrow U(\vec{r})$) and illustrate 
force-induced denaturation for the homopolymer duplex discussed in 
Sec. 2. 
Because $\vec{F}$ appears as a constant {\it imaginary} ``vector potential,'' Eq. (21) 
represents a non-Hermitian generalization of Schroedinger's equation. 
To establish the existence of a first-order transition, we 
write the  solutions of (20) using the right and left
eigenfunctions $\{\phi_{nR}^F(\vec{r})\}$ and 
$\{\phi_{nL}^F(\vec{r})\}$ (with  
a set of common eigenvalues $\{E_n(F)\}$) of $\hat{H}(\vec{F})$,
\begin{equation}
Z(\vec{0},\vec{r}; N)=
\sum_n \phi_{nL}^F(\vec{0})
\phi_{nR}^F(\vec{r})e^
{-E_n(F)Nd/T}.
\end{equation}
We again
invoke ground-state dominance and focus on the behavior of the lowest eigenvalue,
which determines the duplex free energy per unit length $a(F)$ via
\begin{equation}
a(F)=-T \lim_{N\rightarrow \infty} 
\frac{1}{Nd} \ln 
[Z(N,F)],
\end{equation}
where $Z(N,F)$ is defined by integrating over $\vec{r}$,
\begin{equation}
Z(N,F)=\int d^3rZ(\vec{0},\vec{r};N),
\end{equation}
since we work at constant force instead of constant extension.

For $F$ less than a critical value $F_c$ (determined below), we can obtain
the (right) eigenfunctions $\phi_{nR}^{\vec{F}}(\vec{r})$ for $\vec{F}\not=0$ 
from localized eigenfunctions $\phi_n(\vec{r})$ for the Hermitian case 
$\vec{F}=0$ via a ``gauge transformation'' \cite{hatano}.
Indeed, it is easy to check that
\begin{equation}
\phi_{nR}^{\vec{F}}(\vec{r})=
e^{\frac{\vec{F}\cdot\vec{r}}{T}}
\phi_n(\vec{r})
\end{equation}
solves
\begin{equation}
\hat{H}(\vec{F})
\phi_{nR}^{\vec{F}}(\vec{r})=
E_n\phi_n^F(\vec{r})
\end{equation}
(provided $\phi_n(\vec{r})$ satisfies $\hat{H}(\vec{0})
\phi_n(\vec{r})=\epsilon_n\phi_n(\vec{r})$) with the 
{\it same} eigenvalue, $E_n=\epsilon_n$. In 
particular, we obtain from Eq. (23) a negative  $\vec{F}$-independent
free energy per unit length $a=\epsilon_0=-|\epsilon_0|$ for small $\vec{F}$. 
If, however, a localized ground-state eigenfunction
for the Hermitian $\vec{F}=0$ problem decays like
$\phi_0(\vec{r})\sim e^{-\kappa_0|\vec{r}|}$
for large $r$, the new eigenfunctions (25) are only 
normalizable provided $F\le F_c(T)=T\kappa_0(T)$.
Note that $\phi_{0R}^{\vec{F}}(\vec{r})$, which gives
the probability of finding an extension $\vec{r}$, \cite{hatano} is displaced in the direction
of $\vec{F}$, as indicated schematically in 
Fig. 4a. To determine what happens for
$F>F_c$, we check for a nodeless {\it extended}
ground-state wavefunction, as indicated in Fig. 4b.
If $\phi_{0R}^F(\vec{r})\rightarrow {\rm const.}>0$ for large $r$
(i.e., the DNA duplex falls apart), and $\lim_{r\rightarrow\infty}
U(\vec{r})=0$, evaluating Eq. (22) in this limit gives 
$\sl \hat{H} (F)\phi_{0R}^F(\vec{r})=
\frac{-F^2}{2g} \phi_{0R}^F(\vec{r})$, so that 
$a(F)=E_0(F)=\frac{-F^2}{2g}$. The two free energies 
$a(F)$ for $F<F_c$ and $F>F_c$ are plotted in 
Fig. 5 \cite{hatano}. At $F_c$ the two curves intersect at a nonzero 
angle, the classic signature of a first-order phase transition. The latent 
heat of this transition might be observable in a solution of many 
DNA duplexes pulled apart by, say, beads of opposite charge in an 
electric field.
Since the first-order transition curve $F_c(T)$ is given by
$-F_c^2/2g=-|\epsilon_0(T)|$, where $\epsilon_0$ is the smallest
eigenvalue of the transfer matrix for $F=0$, we find using 
Eq. (15) that
\begin{equation}
F_c(T)\propto {\rm const.}|T_m-T|
\end{equation}
near the melting transition for this model.
\begin{figure}[h]
\begin{center}
\centerline{\psfig{figure=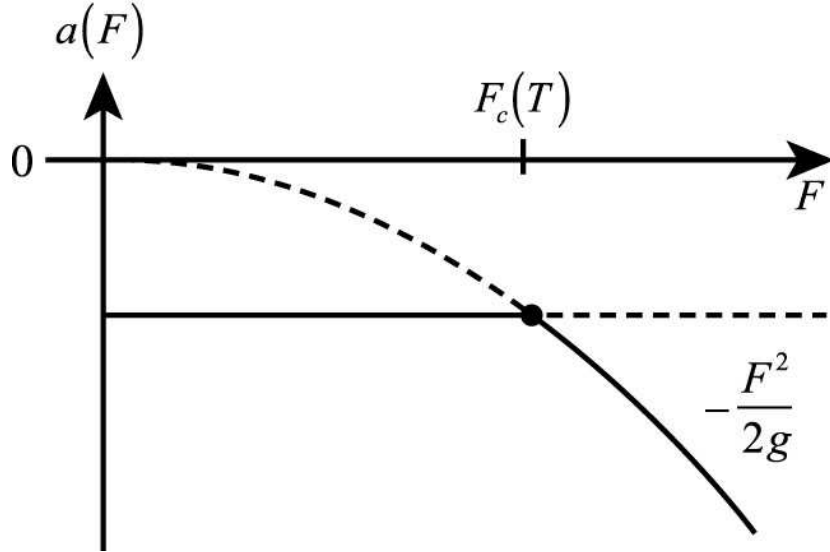}}
\parbox{6.5in}
{\caption{The constant free energy $a=-|\epsilon_0|$ per base of duplex DNA compared to 
the parabolic free energy $a(F)=-F^2/2g$ appropriate for two separated strands. See the 
text for a discussion of more sophisticated freely-jointed chain and 
worm-like chain models of this free energy. The two free energies cross
with discontinuous slope at $F_c(T)$, the location of the first-order
unzipping transition.}
}
\end{center}
\end{figure}

Even though force-induced denaturation is a first-order
transition, there are diverging precursors in the form of DNA
unzipping (see Fig. 1). The probability of an endpoint 
separation $|\vec{r}|$ in the limit $N\rightarrow\infty$ is 
given integrating $|\vec{r}|$ weighted by 
the ground state eigenfunction,
\begin{eqnarray}
\phi_{0R}^{\vec F}
&=
e^{\frac{\vec{F}\cdot\vec{r}}{T}}
\phi_0(\vec{r})\nonumber \\
&\mathop{\simeq}\limits_{r\rightarrow \infty}
e^{\frac{\vec{F}\cdot\vec{r}}{T}}
e^{-\kappa_0|r|}.
\end{eqnarray}
It is then straightforward to show that $\langle|\vec r|\rangle$
diverges as $F\rightarrow F_c^-(T)$ \cite{hatano}
\begin{equation}
\langle|\vec {r}|\rangle
\sim\frac{1}{F_c(T)-F}.
\end{equation}
Upon examining corrections to 
ground state dominance as in the analysis of  
Eq. (19),  we find  that the average number $m$ of
unzipped monomers  diverges, as shown in Fig. 6a,
\begin{eqnarray}
\langle m\rangle &\simeq 
\frac{T}
{d(|\epsilon_0|-F^2/2g)}\nonumber \\
&\sim \frac{1}
{F_c(T)-F}.
\end{eqnarray}
\begin{figure}[ht]
\begin{center}
\centerline{\psfig{figure=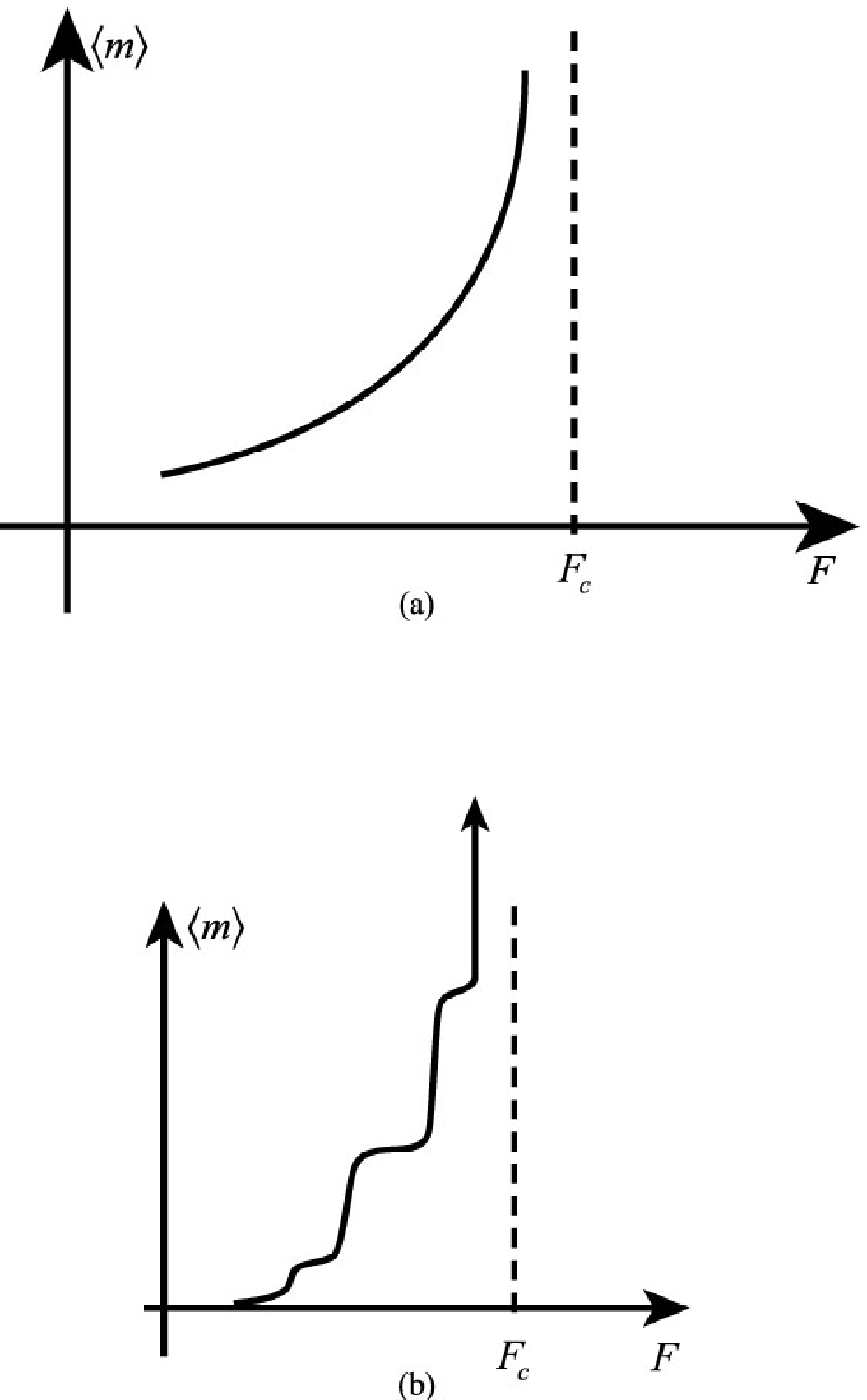}}
\parbox{6.5in}
{\caption{Thermal average $\langle m\rangle$ of the location of the unzipping
fork. (a) Divergence of $\langle m\rangle$ at $F_c$ for a 
homopolymer. (b) Jumps and plateaus in $\langle m\rangle$ (with, on average, a 
much strong divergence) in the presence of sequence disorder.}
}
\end{center}
\end{figure}

The first-order force-induced DNA denaturation transition can be understood
more generally, independent of the mapping onto quantum 
mechanics. If $a_1(T,F)$ is the free energy per unit length
of one of the single-stranded DNA ``handles'' shown in Fig. 7,
and $a_0(T)$ is the (force-independent) free energy per 
unit length of the double-stranded DNA which coexists
in a macroscopically unzipped state, then the condition  for 
two-phase coexistence across the ``meniscus'' or 
unzipping fork at the critical force $F_c$ is \cite{lubensnelson}
\begin{equation}
a_0(T)=2a_1(F_c,T).  
\end{equation} 
The physics of the handle free energy $a_1(F,T)$ is relatively simple, 
provided the force is large enough to neglect self-hybridization and self-avoidance of the 
single-stranded DNA. For example, we could take $a_1(T,F)$ to be the free energy of 
a freely-jointed chain with persistance length $\ell$ \cite{doi} 
\begin{equation} 
a_1(T,F)\propto-\frac{T}{\ell} 
\ln\left(\frac{ T\sinh(F\ell/T)}
{F\ell}\right), 
\end{equation} 
which is proportional to $-\ell F^2/T$ for small $F$. Alternatively one could use a more sophisticated
``worm-line chain'' approximation valid over a larger range 
of forces \cite{markosig}. With an acceptable approximation to $a_1(T,F)$ 
in hand (if necessary, this function could be obtained directly from 
integrating single strand force-extension curves \cite{smith}), one can 
use the experimentally determined phase boundary $F_c(T)$ for 
unzipping in conjunction with Eq. (31) to explore the temperature
dependence of the free energy $a_0(T)$ of double-stranded DNA.
\begin{figure}[h]
\begin{center}
\centerline{\psfig{figure=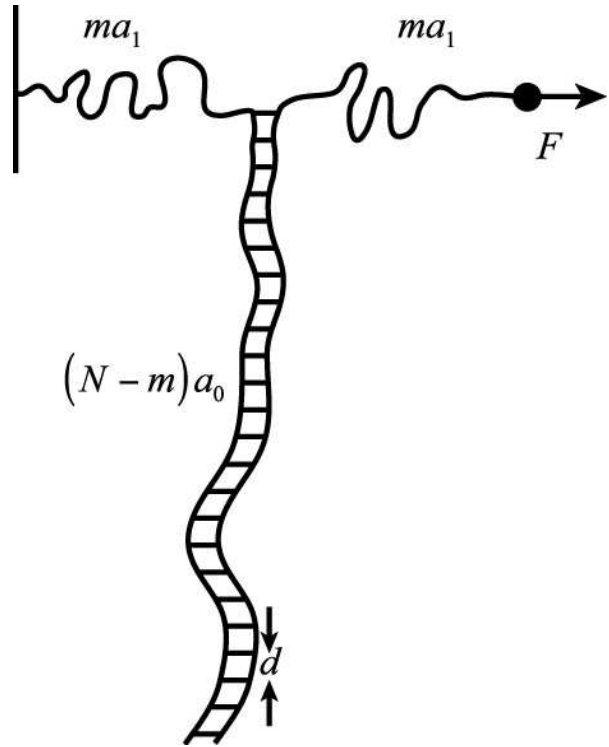}}
\parbox{6.5in}
{\caption{Two-phase coexistence of duplex DNA with free energy 
oer base pair $a_0$ and 
two unzipped ``handles,'' each with free energy per base $a_1$.}
}
\end{center}
\end{figure}

\section{Unzipping with  Sequence Heterogeneity} 

The effect of sequence heterogeneity on thermal denaturation of DNA
at zero force is a subtle and still not completely resolved 
problem, \cite{gros,lyub,tang}. Understanding inhomogeneous base pairing
energies near the first-order force-induced unzipping transition
appears to be more tractable. The integrated effect of 
sequence randomness on the energy landscape for long strands 
of DNA does not change the underlying first-order transition, because it 
only produces corrections of order $\frac{1}{\sqrt{N}}$ 
to the free energies per unit length discussed above, where
$N$ is the total number of base pairs. However, sequence randomness 
has drastic consequences for quantities like $\langle m\rangle$ 
which characterize the unzipping process itself \cite{lubens,lubensnelson}. Here, we simply
sketch the overall picture. Detailed calculations and justifications for 
approximations (such as the neglect of bubbles in DNA) can be 
found in Ref. \cite{lubensnelson}.

Consider the energy $\epsilon(m)$ associated with the conditional 
partition sum of a DNA duplex which has been unzipped by $m$ base pairs.
If the DNA is a homopolymer, consisting exclusively of A:T or G:C base 
pairs, the total energy can be read off from Fig. 7,
\begin{eqnarray}
{\cal E}(m)&=&
2ma_1+(N-m)a_0 \nonumber\\
&=&a_0N+fm
\end{eqnarray}
where
\begin{equation}
f\equiv 2a_1(F)-a_0
\end{equation}
can be expanded to give $f\approx c\frac{2da_1}{dF}|_{_{F_c}}
(F_c-F)$
near the unzipping transition. If we now add sequence randomness, a reasonable
model for the energy of unzipping is
\begin{eqnarray}
\Delta{\cal E}(m)& \equiv &
{\cal E}(m)-Na_0\nonumber \\
& = & fm+\int_0^m dm'\eta (m'),
\end{eqnarray}
where deviations from the average A:T or G:C content are
represented by the function $\eta(m')$ in a convenient 
continuum notation. To the extent that most DNA 
sequences which code for proteins have deviations from the 
average G:C/A:T content describable as a random walk \cite{peng}, we expect 
that the statistical properties of $\eta(m)$ at large distances are those of a 
white-noise random variable,
\begin{equation}
\overline{\eta(m)\eta(m')}
=\Delta_0\delta(m-m')
\end{equation}
where the overbar represents a ``disorder average'' over all possible 
sequences of base pairs. Because G:C and A:T pairing energies 
differ by an amount of order $T$, $\Delta_0\simeq T^2$, where 
we have used the temperature $T
\simeq$ 293--310$^\circ$K characteristic of  most experiments to 
characterize this quantity. The partition function
associated with this simple one-dimensional model involves 
integrating over all possible unzipping lengths $m$, 
\begin{equation}
{\cal Z}(f)=\int_0^\infty dme^{-\Delta{\cal E}(m)/T}.
\end{equation}

The unzipping energy in Eq. (35) is plotted for a particular
base pair sequence (with $\Delta_0=T^2$) in Fig. 8, with $f/T=0.01$. The 
straight line is the energy of a \emph{homopolymer} duplex with 
$\Delta_0=0$ and 
\begin{equation}
\Delta{\cal E}(m)=fm.
\end{equation}
In this case we find immediately from
Eq. (37)  that the thermally averaged degree of unzipping is
\begin{eqnarray}
\langle m\rangle &=&
-T\frac{d\ln {\cal Z}(f)}
{df}\nonumber \\
&=& \frac{T}{f}\sim\frac{1}
{(F_c-F)},
\end{eqnarray}
in agreement with Eq. (30). There are large fluctuations about this average
value, as one can check by evaluating 
\begin{equation}
\sqrt{\langle(m-\langle m\rangle)^2\rangle}=T/f=\langle m\rangle.
\end{equation}
\begin{figure}[h]
\begin{center}
\centerline{\psfig{figure=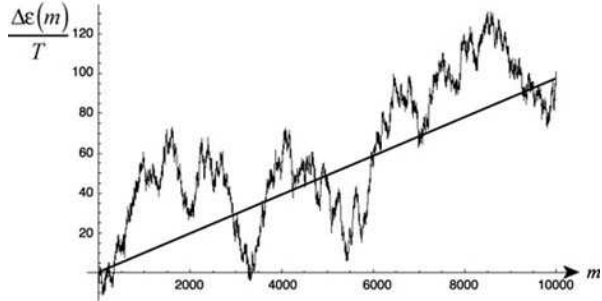}}
{\caption{Energy landscape for 
$F\leq F_c$
for a 
homopolymer duplex (straight line)
and with the addition of inhomogeneous base
pairing energies due to a particular DNA
sequence (jagged line). Here we take 
$\Delta_0 = 
T^2$ and $f/t=0.01$.
}}
\end{center}
\end{figure}
The rugged landscape in Fig. 8 for $\Delta_0\not= 0$ is more 
interesting. Although the positive background slope determined by $f$
insures that the DNA remains macroscopically unzipped, the integrated 
effect of sequence randomness produces deviations from a straight line
which scale like $\sqrt{m}$. These fluctuations lead to deep minima at 
nonzero $m$ with large energy barriers in between. 
A downward fluctuation in this
``integrated random walk'' energy landscape at position $m$ corresponds 
to an energy of order 
\begin{equation}
\Delta\epsilon(m)\simeq fm-
\sqrt{\Delta_0 m}.
\end{equation}
Upon minimizing over $m$, we obtain the estimate
\begin{equation}
\overline{\langle m\rangle}\sim
\frac{\Delta_0}{f^2}\sim \frac{1}{(F_c-F)^2}
\end{equation}
a result confirmed by more elaborate calculations 
\cite{lubens,lubensnelson}. Here, the overbar 
represents an average over a quenched random distribution of DNA sequences.
It can also be shown that thermal fluctuations about this average are more
constrained than for homopolymers, in the sense that
\begin{equation}
\sqrt{
\overline{\langle(m-\langle m\rangle)^2\rangle}
}
\sim \frac{1}{f^{3/2}}\sim
\overline{\langle m\rangle}^{3/4}.
\end{equation}
Thus, in contrast to homopolymers, where 
$\langle(m-\langle m\rangle)^2\rangle / \langle m\rangle^2
={\cal O}(1)$ we now have
\begin{equation}
\overline{\langle(m-\langle m\rangle)^2\rangle}/(\overline{\langle m\rangle}^2
={\cal O}(1/\overline{\langle m\rangle}^{1/2}).
\end{equation}
A typical energy barrier associated with disorder-dominated
unzipping is $f\langle m\rangle+\sqrt{\Delta_0\langle m\rangle}
\sim\frac{\Delta_0}{f}$. Results such as (42) apply when this scale
exceeds $T$, i.e., for $f$ less than a crossover force $f_x$,
\begin{equation}
f<f_x=\frac{\Delta_0}{T}.
\end{equation}
Because $\Delta_0={\cal O}(T^2)$ is large for G:C and A:T base 
pairs, the range of forces where sequence heterogeneity dominates the 
physics is the large region bounded by the left-hand dashed line in 
Fig. 2. Note that disordered dominated prediction for $\overline{\langle m
\rangle}$ in Eq. (42) diverges with a power which is \emph{twice}
as large as the result (39) appropriate for thermally equilibrated 
homopolymers.

Results such as Eq. (42) apply only to a quenched average over an 
entire library of different DNA sequences unzipped in parallel. We can 
understand the behavior for $F\lesssim F_c(T)$ for 
the more
experimentally accessible case of a \emph{particular} 
squence from Fig. 9, which shows the energy landscape for two 
identical sequences with the biases, $f/T=0.01$ and 
$f/T=0=0.006$. The degree of unzipping $\langle m \rangle$ is
dominated by the minimum indicated by an arrow in each case. Somewhere 
between $f=0.01$ and $f=0.006$, the average of $\langle m\rangle$ jumps 
from one minimum to the next. More detailed calculations [14] reveal 
an entire sequence of jumps and plateaus, as illustrated in Fig. 6b. 
Although a best fit to a power law of  this irregular curve should 
reveal the exponent of Eq. (42), the plateaus and jumps themselves 
represent a rough ``fingerprint'' of the individual sequence. 
See Ref. \cite{lubensnelson}, where the 
statistical distribution of plateaus and jumps are
evaluated using the techniques of LeDoussal \emph{et al.} \cite{doussal}.
\begin{figure}[h]
\begin{center}
\centerline{\psfig{figure=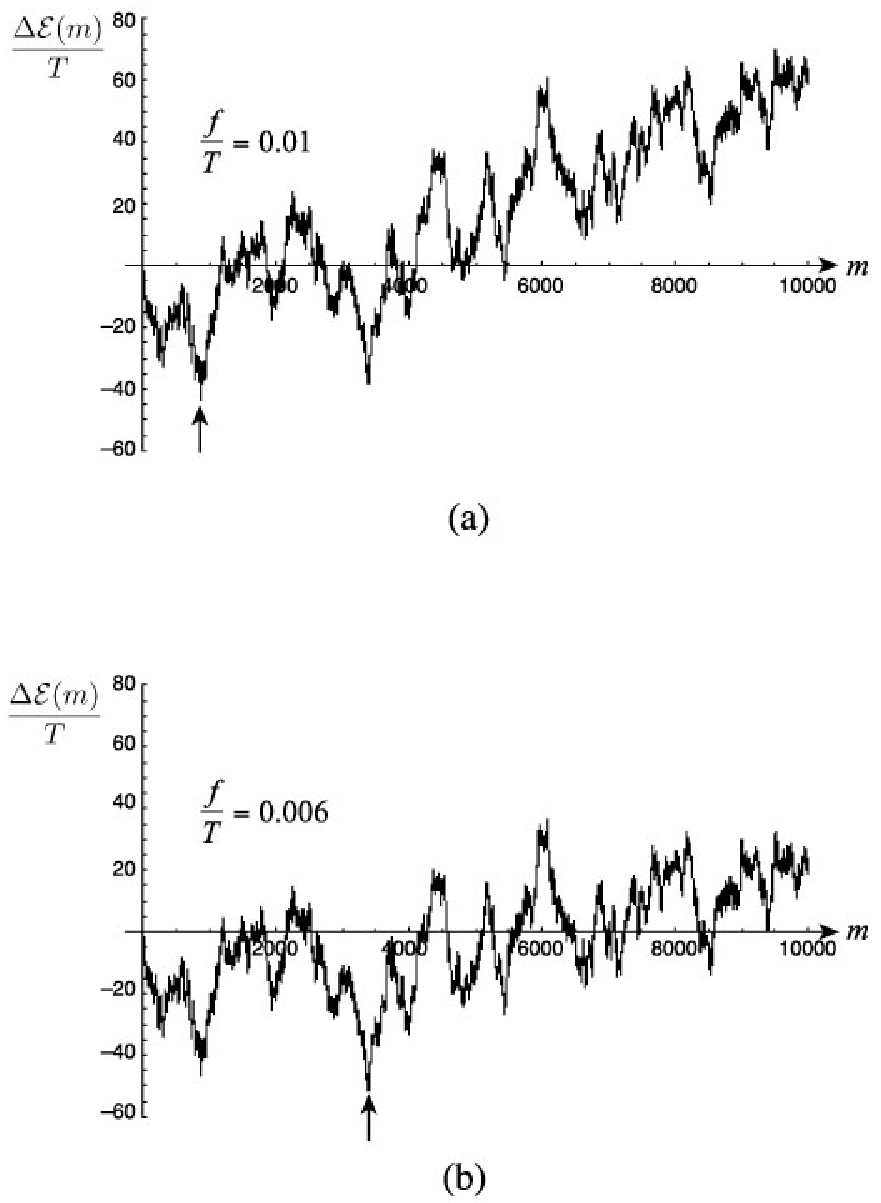}}
\parbox{6.5in}
{\caption{Energy landscapes two identical DNA sequences with 
$\Delta_0=T^2$ and (a) $f/T=0.01$ and
(b) $f/T=0.006$. The arrow indicates the energy minimum  which 
dominates the value of $\langle m\rangle$ in each case.}
}
\end{center}
\end{figure}

\section{Dynamics of Unzipping}

What happens for $F\gtrsim F_c(T)$? In Figure 10 we show an energy 
landscape (for the same sequence as in Fig. 8) with 
$f/T=-0.01$, together with the purely downhill landscape of the 
corresponding homopolymer. Because the average free energy per base pair
of the DNA duplex exceeds the combined free energies per
base of the two single-stranded DNA ``handles'' in Fig. 7, the 
system is   unstable to complete unzipping. It is then appropriate to 
discuss the {\it dynamical} process by which unzipping proceeds on this 
downhill path.
\begin{figure}[h]
\begin{center}
\centerline{\psfig{figure=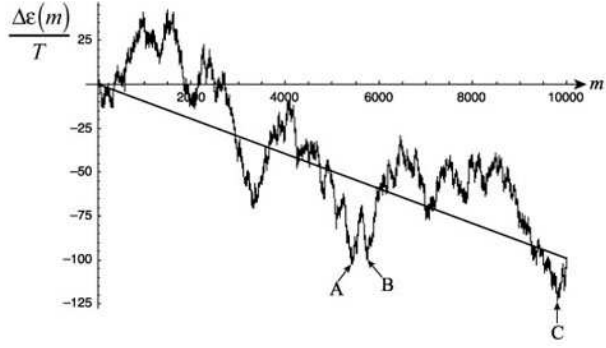}}
\smallskip
\parbox{6.5in}
{\caption{Downhill energy landscape appropriate for
are 
$F\geq F_c(T)$
Although the overall slope is negative, there 
are many traps, such as those at positions A, B, and C. Energy 
barriers between these traps scale as 
$\sqrt{\Delta_0 m}$, where $m$ is 
the number of bases between traps.}}
\end{center}
\end{figure}
Subject to a number of simplifying conditions discussed in Ref. \cite{lubensnelson},
the time dependence of the unzipping fork can be described by an 
overdamped Langevin equation,
\begin{equation}
\frac{dm(t)}{dt}=
-\Gamma\frac{d{\cal E}(m)}
{dm} +\zeta(t),
\end{equation}
where $\Gamma$ is a damping constant $(\Gamma=1/(\tau T)$, where $\tau$
is a microscopic relaxation time) and the noise correlations are
\begin{equation}
\langle\zeta(t)\zeta(t')\rangle=
2T\Gamma\delta(t-t').
\end{equation}
Upon substituting for the energy landscape ${\cal E}(m)$ from Eq. (35), we consider
the case $f<0$ and obtain
\begin{equation}
\frac{dm(t)}
{dt}=\Gamma|f|-\Gamma\eta(m)+\zeta(t),
\end{equation}
which is the equation of motion for a particle with coordinate $m(t)$
executing one-dimensional biased diffusion in a random force field 
proportional to $\eta(m)$.

A great deal is known about this problem \cite{sinai,derrida,bouch}. In the absence of 
sequence heterogeneity ($\Delta_0=0$ in Eq. (36)),  $m(t)$ exhibits diffusion 
with drift at long times, i.e., 
\begin{equation}
\lim_{t\rightarrow\infty}
\frac{\langle m(t)\rangle}{t}=v
\end{equation}
\begin{equation}
\lim_{t\rightarrow\infty}
\frac{\langle[m(t)-\langle m(0)\rangle]^2\rangle
}{t}=2D
\end{equation}
with a well-defined drift velocity $v=\Gamma|f|$ and diffusion constant $D=\Gamma T$. This
is the expected behavior for the straight line homopolymer energy landscape shown in Fig. 10.
Sequence heterogeneity, however, has a dramatic effect on the dynamics unless $|f|$ 
is large. Indeed, the integrated random walk landscape in Fig. 10 has many deep 
minima, even though the average slope is negative. When $|f|$ is small, typical
energy barriers to travel a distance $m$ scale like $\sqrt{\Delta_0 m}$, as is 
reflected in the small barrier connecting the minima labelled A and B 
and the much larger barrier between these minima and the more distant minimum C.
A detailed analysis \cite{sinai,derrida,bouch} reveals three types of anomalous dynamics,
depending on the dimensionless parameter
\begin{equation}
\mu = \frac{2T|f|}
{\Delta_0}.
\end{equation}
\begin{figure}[h]
\begin{center}
\centerline{\psfig{figure=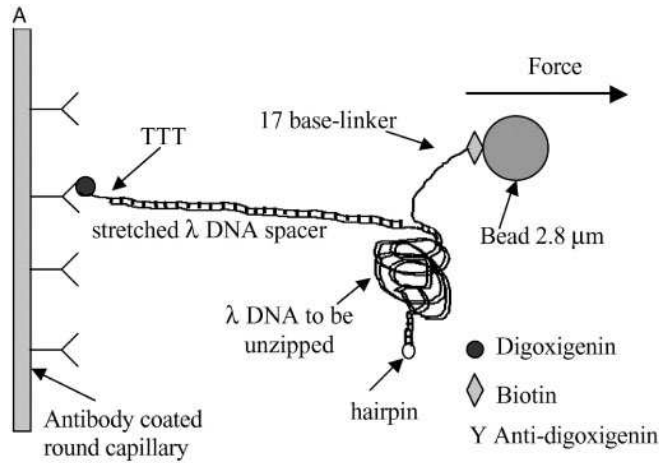}}
\parbox{6.5in}
{\caption{Experimental setup for the force DNA unzipping experiments of 
Danilowicz \emph{et al.} [42]. The unzipping trajectories of dozens of 
genetically identical lambda phage DNAs attached to a thin capillary
tube are tracked. A constant force is generated by attaching magnetic 
beads and placing the apparatus in a magnetic field gradient.}
}
\end{center}
\end{figure}

Right at the unzipping transition, $\mu=|f|=0$ and 
the unzipping fork wanders sub-diffusively according to \cite{derrida}
\begin{equation}
\overline{\langle|m(t)-m(0)|^2\rangle}
\mathop{\simeq}\limits_{t\rightarrow\infty}
\frac{T^4}{\Delta_0^2}\ln^4(t/\tau),
\end{equation}
because of trapping effects arising from  the large energy barriers. 
The conventional diffusion constant $D$, as defined by Eq. (50), vanishes. When
$0<\mu<1$, the unzipping fork drifts downhill, but does so sublinearly with time,
\begin{equation}
\overline{\langle m(t)\rangle}\mathop{\simeq}\limits_{t\rightarrow\infty}
 {\rm const.}\;\;t^\mu,
\end{equation}
so the usual drift velocity $v$ defined by (49) vanishes, with an 
additional anomaly in the spread about this average drift \cite{derrida,bouch}. When this 
bias is large enough so that $1<\mu<2$, one recovers a
well-defined drift velocity $v$, but with 
superdiffusive spreading,
\begin{equation}
\overline{\langle[m(t)-\langle m(t)\rangle]^2\rangle}
\mathop{\sim}\limits_{t\rightarrow\infty}
t^{2/\mu}.
\end{equation}
Conventional diffusion with drift is only recovered for forces large enough so that $\mu>2$, 
i.e., for
\begin{equation}
-f>f'_x=
\frac{\Delta_0}{T}.
\end{equation}
This critical force is the same order of magnitude as the crossover scale defined {\it below} the 
unzipping transition by Eq. (45) and leads to the right-hand dashed line in Fig. 2.
The parameters of DNA are such that anomalous unzipping appears in a very large range 
(compared to typical disorder effects in conventional critical phenoman \cite{harris}), 
roughly $\frac{1}{2} F_c\lesssim F \le \frac{3}{2} F_c$.
\begin{figure}[p]
\begin{center}
\centerline{\psfig{figure=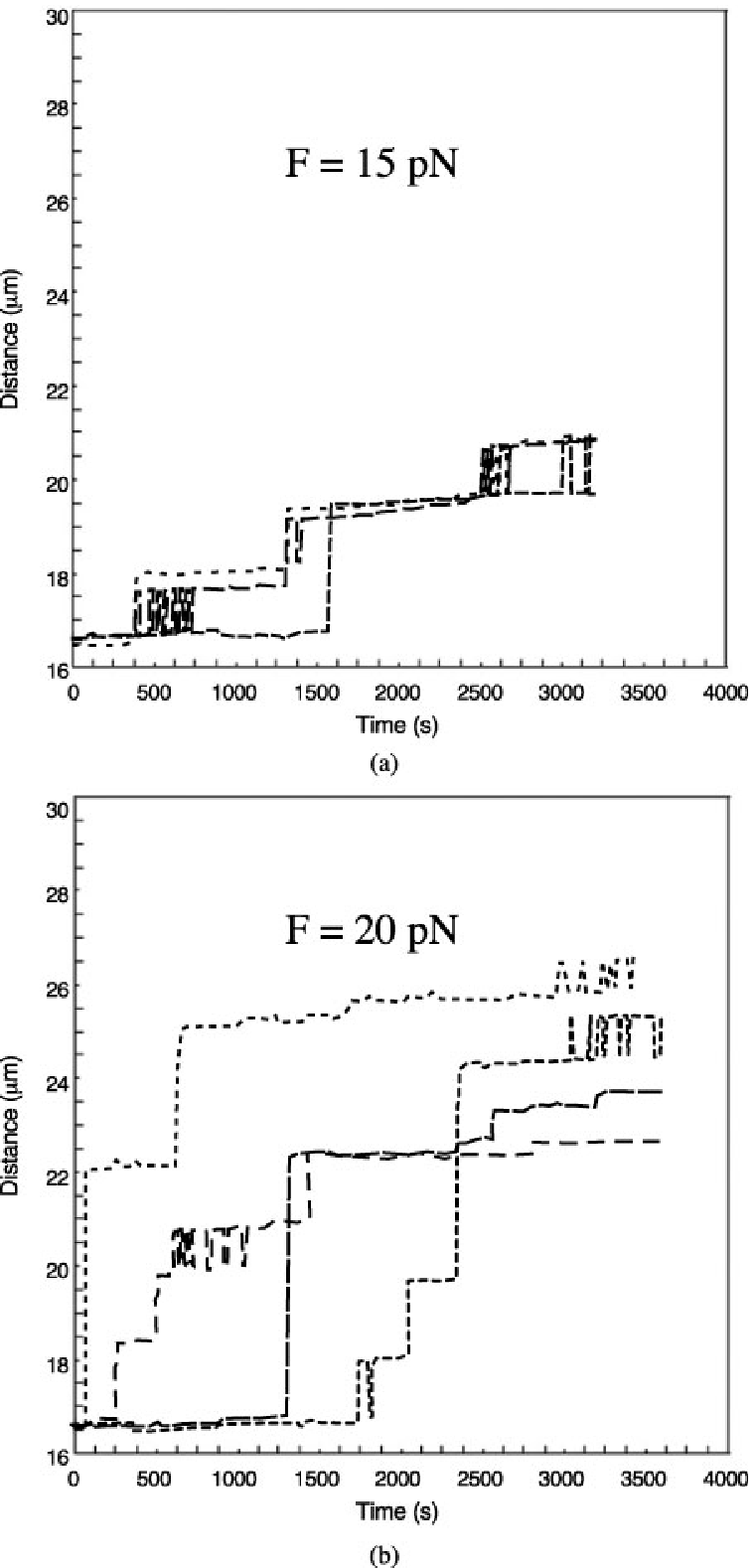}}
\parbox{6.5in}
{\caption{Unzipping histories obtained from the 
apparatus of Fig. 11 for (a) three beads with $F=$15 $pN$ and (b) four 
beads with $F=$20 $pN$. Note the long-time scale (of order 1 hour)
and the many jumps and plateaus. The oscillations arise when the 
unzipping fork encounters two nearly degenerate energy minima. 
The same plateaus recur for different 
beads at the same force, suggesting that these are due to 
sequence disorder (from Ref. [42]).}
}
\end{center}
\end{figure}
\section{Pauses, Jumps and Concluding Remarks}

Experiments which study the mechanical denaturation 
of lambda phage  DNA at constant force have recently been carried out in the 
group of Mara Prentiss [42]. Digoxigenin linkers attach several dozen identical 
copies of this DNA along a long thin capillary tube. Each DNA
to be unzipped  is  offset by an additional
DNA spacer (see Fig. 11). Biotin links the other end of the DNA to  magnetic 
beads. Unzipping at constant force
is achieved by applying  a uniform magnetic field gradient, causing 
the beads to stretch out at right angles to the capillary tube, like a 
series of flags on a flagpole.

Figure 12a records some unzipping histories at $T=20^\circ$C for three
different beads at different heights on this ``flagpole'' with 
$F=15$~$pN$, corresponding approximately to the critical unzipping force 
for the front half of the sequence for phage lambda. 
(The back half of this DNA has a lower G:C content.) The beads are well separated, and are 
clearly subject to microscopically different thermal histories. Nevertheless,
there are distinctive pauses (some lasting 15 minutes or more) and jumps
common to the different beads. As the individual DNA's unzip, the beads
occasionally jump back and forth, suggesting two nearly degenerate minima in the 
energy landscape about 1000 base pairs apart. This behavior is similar
to that found for much shorter RNA hairpains in Ref. \cite{lip}. The long pauses
before unzipping resumes are consistent with large energy barriers, of
order 20--30~$T$. In contrast, unzipping at a slow constant 
velocity is an {\it imperative} in the work of Bockelmann {\it et al.} \cite{bock,esse,bocktho}.
Because these experiments are {\it effectively} at constant extension, the force in 
this case fluctuates as needed about $F_c(T)$ to insure that all
energy barriers are overcome.

Many of the pauses and local ``two-level system'' oscillations found by 
the Prentiss group coincide with those predicted for phage lambda DNA \cite{danil}, 
using energy landscapes obtained from the base pairing and stacking 
energies of Santa Lucia {\it et al.} \cite{lucia}. Overall, these experiments  seem 
consistent with important aspects of the theory reviewed here, such 
as large energy barriers and sequence-specific pause points. 
We should stress that longer run times or higher temperatures are 
necessary to obtain true thermodynamic equilibrium. Figure 12b shows 
the unzipping history of four different beads at $F\simeq 20$~$pN$. There 
are again jumps between common plateaus as well as oscillations 
suggesting nearly degenerate minima. Although 20 $pN$ should be above the 
thermodynamic critical force $F_c(T)$, the DNA nevertheless fails to 
unzip completely on experimental time scales, presumably due to the ruggedness of ``downhill'' 
landscapes such as that in Fig. 10. Predictions such as Eq. (42) are 
only be applicable to systems in true thermodynamic equilibrium.

Experimental checks of phase diagrams such as that in Fig. 2 and the 
associated physical predictions would be of considerable interest. As
mentioned in Sec. 3, the phase diagram itself can be used to directly
measure the temperature-dependent free energy of the DNA duplex at zero
force. Analysis of this free energy is simplified if a theory of 
the unzipped handles is available. Analytic theories require forces large enough to 
stretch out the handles and prevent hairpins created by self-hybridization.
The self-hybridization of ssDNA (for $F\lesssim 10$ $pN$) indicated in the 
experiments of Maier {\it et al.} \cite{maier} could be reduced by 
introducing single stranded  binding proteins to smooth out the heterogeneity of the 
handles.

The assumption (36) of short-range correlations in the sequence is an 
approximation valid only for simple ``coding DNA'' in bacteria or viruses. 
The introns or so-called
``junk DNA'' present in many eukaryotic organisms may be  better
described by \cite{peng} 
\begin{equation}
\overline{ \eta(m)\eta(m')}
\sim\frac{1}
{|m-m'|^{2-2\beta}},
\end{equation}
with $\beta>1/2$. Although we are unaware of detailed
calculations, a straightforward generalization of the argument
leading to Eq. (42) now gives \cite{lubens}
\begin{equation}
\overline{\langle m\rangle}\sim
1/f^{\frac{1}{1-\beta}}.
\end{equation}
For $\beta\le 1/2$, we expect that predictions for 
sequences with short-range correlations will apply. The 
exponent $\beta=0.55$ seems to describe the energy 
correlations of some  intron-rich  sequences [36].

Ideas similar to those discussed here
can also be applied to force-induced denaturation \cite{gerland,lubensnelson} of the 
multiple stem-loop structures which define the secondary structure of 
complicated RNA enzymes. Here, however, strings of RNA hairpins are 
unzipped simultaneously and the discrete jumps and 
plateaus in $\langle m \rangle$ near the unzipping force are averaged out.
\begin{acknowledgments}
Virtually all the theoretical work described here is contained in 
the Ph.D. thesis of my former student, David Lubensky. I am 
fortunate to have such students. I am grateful as well to
Claudia Danilowicz, Mara Prentiss and other experimenters for 
their determination to test some of the predictions sketched 
here. Finally, I would like to thank D. Branton, R. Bundschuh, T. Hwa,  
and Y. Kafri for helpful conversations on this material.
This work was supported by the National Science Foundation, 
through grant DMR 0231631 and in part through the 
Harvard Materials Research Laboratory, via Grant DMR 0213805.
\end{acknowledgments}

\end{document}